\documentstyle[twoside,fleqn,espcrc2,epsf]{article}

\newcommand{\ee}{\end{equation}}
\newcommand{\be}{\begin{equation}}
\newcommand{\bea}{\begin{eqnarray}} 
\newcommand{\eea}{\end{eqnarray}}

\title{Calculation of the renormalized charmed-quark mass in Lattice 
QCD\thanks{Talk presented by A. Bochkarev at LATTICE96}}
 
\author{A. Bochkarev
and Ph. de Forcrand\\[0.1cm]
 Swiss Center for Scientific Computing, ETH-Zentrum, 
CH-8092 Z\"urich, Switzerland  }
\begin{document}
\begin{abstract}
The correlator of heavy-quark currents is calculated in quenched 
Lattice QCD on a $16^3 \times 32$-lattice for $\beta= \{ 6, 6.3 \}$.
The renormalized charmed quark mass is extracted from the short-distance 
part of that correlator: $m_c^{\overline{MS}}(m_c)\,=\,1.22(5)\,GeV$.
We study the sensitivity of our data to the strong coupling constant.
\end{abstract}

\maketitle

The point-to-point correlator of local heavy-quark currents:
\be
\Pi(q^2)_{\mu\nu}\;=\;i\int dx e^{iqx} <0| T\left\{j_{\mu}(x) 
j_{\nu}(0) \right\}|0>                                        \label{p(q)}
\ee
where $j_{\mu} \,=\, \bar{c}\,\gamma_{\mu}\,c$, is reliably calculable in 
the loop expansion of perturbative QCD in the vicinity of zero momentum. 
If one knows the correlator from experiment or
from lattice simulations one can then extract parameters of
perturbative QCD such as the renormalized heavy-quark mass or strong
coupling constant normalized on the heavy-quarks threshold \cite{svz}.
In earlier work \cite{b,bdf} we pointed out that 
applicability of the perturbative loop-expansion implies that the lattice 
artifacts of the short-distance part of the correlator $\Pi(q^2 \sim 0)$ 
may be studied analytically order-by-order in that expansion. Using this idea,
our aim is to extract the renormalized charmed-quark mass $m_c$ and strong 
coupling constant $\alpha_{s}(m_c)$.

Following \cite{svz} we consider the ratios 
$\,r_{n}\;=\;{\cal M}_{n+1}/{\cal M}_{n} \,$ of moments of the correlator
(\ref{p(q)}):
\be
{\cal M}_{n}\;=\; \frac{1}{2^{2n}\, n! (n+1)!}\,
\int d^4x \, x^{2n} \,\Pi(x)                        \label{xmoment}
\ee
to separate the short and the long distances.
The applicability of perturbative QCD near $q^2 = 0$ implies the
following expansion for the {\em lower} ratios $r_{2,3,4}$, 
originating from short-distances:
\begin{equation}
r_{n}\;=\; \frac{a_n}{4 m_{c}^{2}} \,\left\{ \;1\;+\; \omega_n\, 
\alpha_{s}(m_c)\; \right\}                        \label{rqcd}
\end{equation}
where $\{a_n, \omega_n\}$ are known numbers \cite{svz}.  
The term $\propto a_n$ comes from one loop of free charmed quarks.
The term $\propto \omega_n$ comes from the two-loop diagrams due to
one-gluon exchange. 

\begin{figure}[htb]
\vspace{-.9cm}
\epsfxsize=65truemm
\centerline{\epsffile[25 150 490 700]{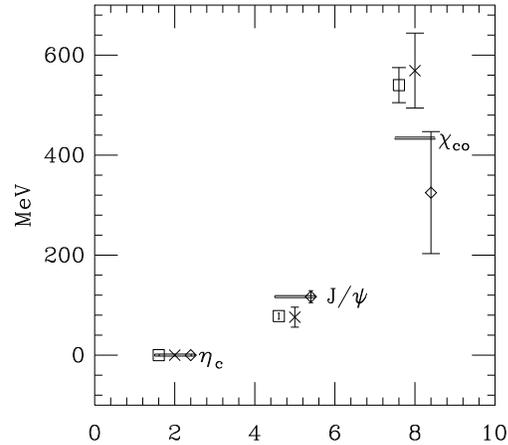}}
\vspace{-1.5cm} 
\caption{{\small Charmonium spectrum obtained on a $16^3\times 32$ 
lattice at $\beta = 6$ (squares) and $\beta = 6.3$ (diamonds) and on 
an $8^3\times 16$ lattice at $\beta = 6$ (crosses) with the 
clover-and-tadpole-improved action. The pseudoscalar mass 
$m_{\eta_{c}} = 2.979GeV$ is used for normalization. 
The wide horizontal lines are experimental data.
\label{fig:spec} } }
\vspace{-1cm}
\end{figure}

We have calculated the correlator (\ref{p(q)}) in quenched QCD on 
an $8^3\times 16$ lattice for $\beta = 6$ and a $16^3\times 32$ lattice
for $\beta = \{ 6, 6.3 \}$ with the Wilson tadpole-and-clover-improved fermionic
action \cite{action}, with $20$ configurations in every case.
We fixed the scale (lattice spacing $a$) from the mass of the low-lying
resonance. To extract that mass in these relatively small volumes we
incorporated the physical continuum spectrum at high energies in the
way it is done in the {\em QCD sum rules} approach \cite{svz} to 
describe the behavior of the correlator at short distances. 
We used the following extension of the dispersion relation 
for the two-point correlator $\Pi (q^2)$ to the lattice theory \cite{bdf}:
\be
\Pi (q^{2})\;\;=\;\; \int ds\, \frac{\rho(s)}{s\,+\,\frac{4}{a^2}
\sum_{\mu}^{4} \sin^{2}(q_{\mu} a/2)}                      \label{disp}
\ee
The {\em higher} ratios $r_{8}\,-\,r_{11}$ originate from long distances.
They are saturated by the single low-lying resonance contribution.
Our results for the charmonium spectrum are shown in Fig. 1.

\begin{figure}[htb]
\vspace{-1.1cm}
\epsfxsize=65truemm
\centerline{\epsffile[25 150 490 700]{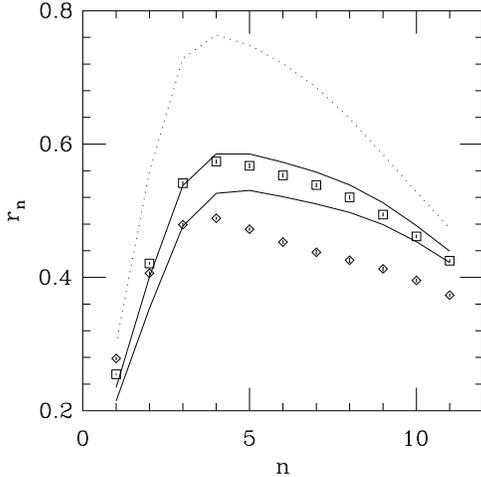}}
\vspace{-1.5cm}
\caption{{\small Monte-Carlo data for the vector currents on a 
$16^3 \times 32$ lattice at $\beta = 6$
(diamonds), $\beta = 15$ (squares); $\kappa = 0.1100$.
The lines show the ratios of moments for one loop of free Wilson 
fermions. The dashed line corresponds to $\kappa = 0.1100$;
the solid lines fit the $3^{rd}$ ratio of Monte-Carlo data by construction.
\label{fig:beta15} } }
\vspace{-1cm}
\end{figure}

We fit the {\em lower} ratios $r_{3,4}$ of Monte-Carlo data with the
free-quarks approximation to the correlator (\ref{p(q)}):
\begin{equation}
r_{3, 4}(\mbox{free Wilson quarks})\,=\,r_{3, 4}(\mbox{Monte-Carlo}) \label{mc_approx}
\end{equation}
The Wilson parameter $\kappa $ corresponding to those free quarks 
determines the renormalized quark mass. The relation between $\kappa$ in
a finite volume and the corresponding quark mass of the continuum theory
was clarified in \cite{bdf}. 

To specify the subtraction scheme of the renormalized quark mass, one 
should be sensitive to the $\alpha_s$-correction in the ratios $r_n$.
If we ignore lattice artifacts in the coefficients $\omega_n$ and take
$\omega_n$ from the continuum theory \cite{svz}, choosing the 
${\overline{MS}}$ subtraction scheme and $\alpha_s(m_c) \approx 0.3$ \cite{bdf},
we get the results for the renormalized
charmed-quark mass shown in Table 1. Since the two-loop correction is 
small in the {\em lower} moments, the uncertainty in the quark mass  
introduced by taking $\omega_n$ from the continuum theory rather than 
evaluating it on the lattice is small.

\begin{table}[htb]
\vspace{-.6cm}
\caption{{\small The charm-quark mass obtained by fitting the ratios 
$r_3$, $r_4$ of the correlator on the $16^3 \times 32$ 
lattice 
. 
The masses $m_{res}$ and $m_{c}$ are in units of the lattice 
spacing. The mass $m_{c}^{\overline{MS}}[a_{J/\psi}]$ is in $GeV$ with 
the lattice spacing fixed from the vector channel. 
Errors are statistical.   }    }
\begin{center}
\begin{tabular}{|c|c|c|c|}
\hline
\multicolumn{4}{|c|}{$\,\beta = 6.0 ,\;\;\;\,\kappa=0.1060,
\;\;\;(\,a\approx\,1.9 \,GeV)$}                             \\
\hline
fit to $r_3$&$ m_{res} $&$ m_{c} $&$
 m_{c}^{\overline{MS}}[a_{J/\psi}]$ \\ \hline
$\eta_c$ & 1.562(5) & .618(2) & 1.208(6) \\
$J/\psi$ & 1.600(6) & .660(1) & 1.246(6) \\
$\chi_o$ & 1.86(4)  & .645(2) & 1.230(8) \\ \hline
fit to $r_4$&$ m_{res} $&$ m_{c} $&$ 
m_{c}^{\overline{MS}}[a_{J/\psi}]$ \\ \hline
$\eta_c$ & 1.562(5) & .645(1)  & 1.224(6) \\
$J/\psi$ & 1.600(6) & .677(1)  & 1.258(6) \\
$\chi_o$ & 1.86(4)  & .686(3)  & 1.27(1)  \\ \hline
\end{tabular}
\begin{tabular}{|c|c|c|c|}
\hline
\multicolumn{4}{|c|}{$\,\beta = 6.3 ,\;\;\;\,\kappa=0.1150,
\;\;\;(\,a\approx\,3.7 \,GeV)$}                            \\
\hline
fit to $r_3$&$ m_{res} $&$ m_{c} $&$
 m_{c}^{\overline{MS}}[a_{J/\psi}]$\\ \hline
$\eta_c$ & 0.805(6) & .307(2)  & 1.149(6) \\
$J/\psi$ & 0.836(7) & .336(2)  & 1.214(9) \\
$\chi_o$ & 0.89(6)  & .338(5)  & 1.23(3)  \\ \hline
fit to $r_4$&$ m_{res} $&$ m_{c} $&$
m_{c}^{\overline{MS}}[a_{J/\psi}]$ \\ \hline
$\eta_c$ & 0.805(6) & .321(2)  & 1.165(6) \\
$J/\psi$ & 0.836(7) & .344(2)  & 1.222(6) \\
$\chi_o$ & 0.89(6)  & .359(8)  & 1.27(4)  \\ \hline 
\end{tabular}
\end{center}  
\vspace{-.8cm}
\end{table}

One can see a rather stable value of the renormalized charmed-quark mass 
$m_c^{\overline{MS}}(m_c)\,=\,1.22(5) GeV$, which is in good agreement with 
estimates of the continuum theory \cite{nar}: 
$m_c^{\overline{MS}}(m_c)\,\approx \,1.23 GeV$. 
The previously reported lattice result is \cite{mass} : 
$m_c^{\overline{MS}}(m_c)\,=\,1.5(3) GeV$.

\begin{figure}[htb]
\vspace{-1cm}
\epsfxsize=65truemm
\centerline{\epsffile[25 150 490 700]{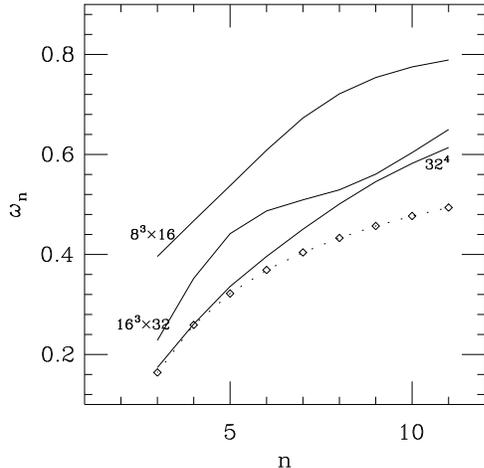}}
\vspace{-1.5cm}
\caption{{\small The coefficient $\omega_n$ obtained via the lattice 
dispersion relation with the Schwinger spectral density on different 
lattices for $a m_c \,=\,0.6$ vs. the continuum-theory result (dashed line). 
} }
\label{fig:omega}
\vspace{-.8cm}
\end{figure}

We have found previously \cite{bdf} that the lattice version (\ref{disp}) 
of the dispersion relation works extremely well $(i)$ to reproduce the lattice  
correlator on the one-loop level in perturbation theory, and $(ii)$ to fit
Monte-Carlo data in small volumes with the phenomenological form of
spectrum incorporating both the low-lying resonance and the smooth
continuum spectrum at high energies.
Here we use the dispersion relation (\ref{disp}) to estimate lattice 
artifacts in the coefficients $\omega_n$. We use in eqn.(\ref{disp}) 
the Schwinger expression for the spectral density, which determines the 
correlator of vector currents on the two-loop level. 
The quark mass is fixed 
in the $\overline{MS}$-scheme in the continuum theory (see \cite{svz}, where
it was fixed in the $MOM$-scheme). One can see from Fig.3 that the 
coefficient $\omega_n$ obtained in this way remarkably coincides with the 
continuum-theory results for the ratios of interest $n = 3, 4$ already on a 
$32^4$ lattice. Since the use of the lattice 
dispersion relation (\ref{disp}) helps to approach the continuum limit
faster \cite{bdf}, Fig.3 does suggest that one can have very small lattice artifacts in the coefficients $\omega_n$ on a $32^4$-lattice for reasonable 
values of the quark mass $a m_c\,\sim\,0.6$. Finite-size effects are seen 
to be significant on smaller lattices.

Keeping $\kappa$ the same as at $\beta=6$,
we have generated Monte-Carlo data on a $16^3\times32$ lattice for 
$\beta=\{ 10, 15\}$. Moment ratios are shown on Fig.2 for $\beta=15$.
The corresponding fermion mass $\tilde{m}_c$ is very heavy, since
$\tilde{m}_c\cdot a[\beta=15]=m_c\cdot a[\beta=6]$. Nonperturbative effects 
are then strongly suppressed in 
the range of distances probed by the given lattices, hence only perturbative 
terms survive. Indeed one can see the Monte-Carlo data much better fitted 
with one loop
of free Wilson fermions in a wide range of ratios at $\beta=15$ than at
$\beta=6$, which is in the confining phase. 
Nonetheless the deviation from the one-loop approximation is clearly visible.
Therefore, we fit our data with the two-loop expression (\ref{rqcd}), taking
 the coefficients $\omega_n$ from continuum theory. 
Two adjacent ratios $r_n$, $r_{n+1}$ can be used to fix the quark mass and
the effective coupling constant $\alpha_s'$.
Fig.4 shows our results for $\alpha_s'$.
Since $\alpha_s' = \alpha_s(\tilde{m}_c) \sim \alpha_s(a)$, 
one can check that
the relative height of the plateaus in $\alpha_s'$ for $n= 5 - 8$
is consistent with asymptotic scaling, as expected at those high $\beta$s.

\begin{figure}[htb]
\vspace{-1cm}
\epsfxsize=65truemm
\centerline{\epsffile[25 150 485 680]{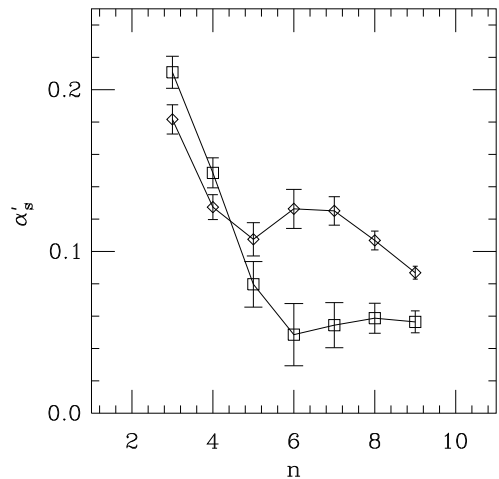}}
\vspace{-1.3cm}
\caption{{\small Strong coupling constant for very heavy fermions obtained 
on a $16^3\times 32$ lattice at $\beta = 15$ (squares) and 
$\beta = 10$ (diamonds) for $\kappa = 0.1100$. } }
\label{fig:als}
\vspace{-1.2cm}
\end{figure}

\end{document}